\PassOptionsToPackage{table}{xcolor}
\documentclass[11pt]{article}

\usepackage[final]{acl}

\usepackage{times}
\usepackage{latexsym}
\usepackage{amssymb}
\usepackage{amsmath}
\usepackage{subcaption}
\usepackage{booktabs}
\usepackage{multirow}
\usepackage{pifont}
\usepackage{array}
\usepackage[T1]{fontenc}
\usepackage[utf8]{inputenc}

\usepackage{microtype}
\usepackage{xurl}
\usepackage{hyperref}

\usepackage{inconsolata}

\usepackage{graphicx}

\usepackage{tabularx}

\usepackage{amsmath}
\usepackage{amsfonts}
\usepackage{nicefrac}

\usepackage{booktabs}
\usepackage{tabularx}
\usepackage{array}
\usepackage{wrapfig}
\usepackage{graphicx}
\graphicspath{{./images/}}

\definecolor{CornellRed}{rgb}{0.7, 0.11, 0.11}
\definecolor{OliveGreen}{rgb}{0,0.6,0}

\definecolor{HeaderBlue}{HTML}{DCE6F1}
\definecolor{SectionGray}{HTML}{F1F3F5}
\definecolor{PersonalizedBlue}{HTML}{EEF5FF}
\definecolor{GenericGold}{HTML}{FFF5E1}
\definecolor{BaselineGray}{HTML}{F7F7F7}
\definecolor{OtherPurple}{HTML}{F5F0FF}
\definecolor{AlternateGray}{HTML}{F1F3F5}
\definecolor{BestGreen}{HTML}{E8F4EA}
\definecolor{CitationBlue}{HTML}{0000FF}
\definecolor{HeaderBlue}{HTML}{DCE7F3}
\definecolor{SectionGray}{HTML}{F0F2F4}
\definecolor{SolLowBg}{HTML}{F3F4F5}
\definecolor{SolHighBg}{HTML}{EBF3FF}
\definecolor{SolMaxBg}{HTML}{F4EEFC}
\definecolor{AstraMedBg}{HTML}{FFF5DF}

\title{Do Personalized Skills Help Coding Agents? An Empirical Study of Developer Interaction Traces} 

\author{%
  Shuyan Huang\\
  UMass Amherst \\
  \texttt{shuang@umass.edu} \\
  \And
    Kai Du \\
  OpenRefinery.ai\\
  \texttt{kai@openrefinery.ai} \\
  \And
  Andrew Lan \\
  UMass Amherst \\
  \texttt{andrewlan@umass.edu} \\
}

\begin{document}
\maketitle
\begin{abstract}
As human developers increasingly collaborate with large language model (LLM)-powered coding agents, adapting to individual developer preferences, potentially via agent skills, may offer way to reduce repeated guidance and improve collaboration efficiency. %Agent skills provide a lightweight mechanism for capturing and reusing such preferences across tasks. 
However, existing approaches primarily focus on building skills for each task rather for each human developer. 
%task-specific knowledge, leaving the effectiveness of developer-specific personalization largely unexplored. 
In this work, we propose a framework that combines rule-based bootstrapping with LLM-based refinement to distill developer-specific skills from their interactions with coding agents. We further develop an interactive replay framework with an LLM-based developer simulator for evaluation. 
%to evaluate their effectiveness on randomly held-out tasks from the same developers. 
We conduct experiments on a collection of replayable real-world developer-agent interaction sessions. 
We find that overall, task-specific skills are the most effective; However, personalized skills become more effective when developer preferences are manifested with enough frequency, especially when developers work on tasks that are highly related. These findings provide empirical insights suggest that i) broadly transferable procedural knowledge can be more robust than developer-specific preference signals and ii) effective personalization requires mining from rich user preference signals.
%Across two skill generators and four coding-agent executors, personalized-skill effectiveness is highly sensitive to the generator--executor pairing. Across multiple skill variants, generic skills perform best, suggesting that broadly shared procedural guidance is more effective than developer-specific personalization. Further analysis suggests that personalized skills become more effective when developer preferences appear frequently, particularly when their traces contain multiple examples relevant to held-out tasks. 
\end{abstract}

\section{Introduction}

Large language model (LLM)-based agents have rapidly evolved from code-completion tools into interactive systems for solving complex software engineering (SWE) tasks, such as navigating repositories and iteratively repairing failed solutions~\citep{yang2024swe,jimenez2024swe,wang2025openhands}. As developers collaborate with these agents, they often have distinct expectations regarding how code changes should be implemented~\citep{baumann2026swechat}. In practice, these preferences are often implicit at first and emerge across repeated interactions with the agent on multiple tasks~\citep{zhao2025llms,baumann2026swechat}. An agent capable of inferring developer preferences from observed interaction trajectories has the potential to adapt its behavior, improving the efficiency of developer-agent collaboration~\citep{wu2024longmemeval,liang2026learning}. For example, a developer may consistently prefer minimal, localized changes over broad refactoring, even when this preference is not explicitly stated in every task. An agent can proactively incorporate these preferences into its thinking and implementation processes when solving subsequent tasks, thereby reducing the need for developers to repeatedly specify the same preferences and provide similar feedback across tasks. 
% \ml{some cites here will make the para look better}

\begin{figure*}[!tpbh]
    \centering
    \includegraphics[width=1\textwidth]{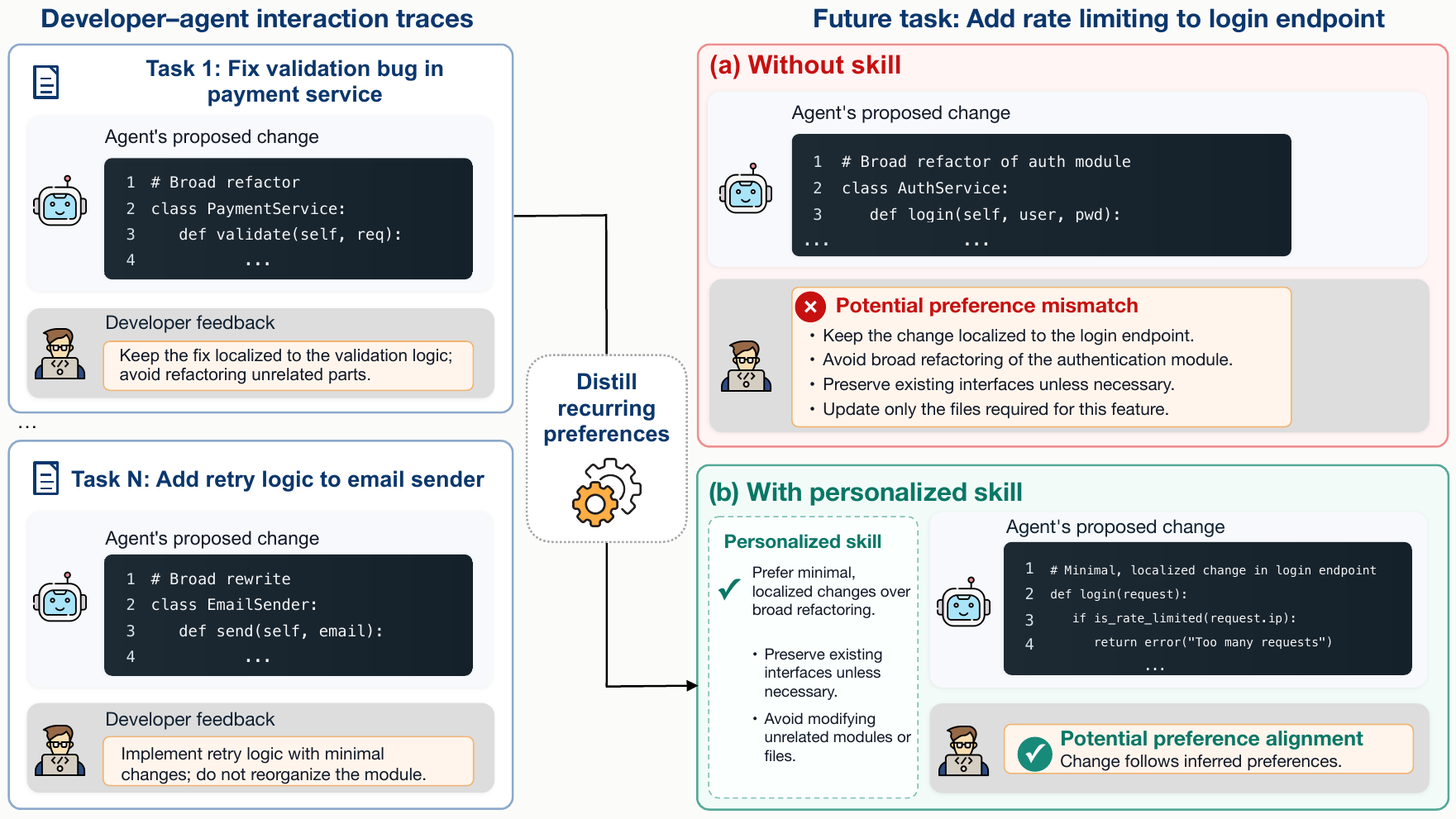}
    \caption{Illustration of how recurring developer preferences are distilled from developer–agent interaction traces into personalized skills and applied to guide coding agents on new tasks.}
    \label{fig:illustraction}
\end{figure*}

Agent skills have recently emerged as a promising mechanism for retaining and transferring experience, by packaging reusable procedural knowledge as structured instructions for LLM-based agents~\citep{han2026swe,xu2026agent}. Current methods distill and optimize the skills from execution trajectories to improve coding-agent performance without updating the underlying model parameters~\citep{ni2026trace2skill,yang2026skillopt,wang2026skillgrad}. These approaches primarily distill domain- or task-specific knowledge to improve agent performance at a specific task. However, in practice, each human developer may work on a diverse collection of tasks that may not have significant overlap with that of other developers. It remains unexplored whether a task-independent, yet developer-specific natural-language skill distilled from a developer's interactions with the coding agent can generalize across their tasks.

To bridge this gap, we investigate whether we can learn personalized agent skills that capture transferable developer-specific preferences rather than task-specific behaviors. As illustrated in Figure~\ref{fig:illustraction}, recurring feedback across a developer's tasks can reveal stable implementation preferences, which can be distilled into reusable guidance and applied to new tasks. However, generating effective personalized skills and evaluating them present two main challenges. First, a developer's observed interaction traces contain a mixture of transferable preferences, task-specific requirements, and one-off corrections, making it challenging to distinguish reusable preferences from task-specific behavior. Moreover, these preferences are often conveyed implicitly through feedback on specific implementation rather than explicit instructions, requiring us to infer the underlying preferences from implicit signals. Second, fairly evaluating personalized skills depends on comparing different skills under identical interaction conditions. This evaluation requires us to faithfully reconstruct the original developer-agent interaction session while isolating the effect of the injected skill. Thus, we investigate the following core research question:

\begin{quote}
\emph{Can we distill personalized skills from a subset of a developer's interactions with coding agents, and do they improve task success and reduce pushback on held-out tasks from the same developer?}
\end{quote}

\begin{figure*}[!tpbh]
    \centering
    \includegraphics[width=1\textwidth]{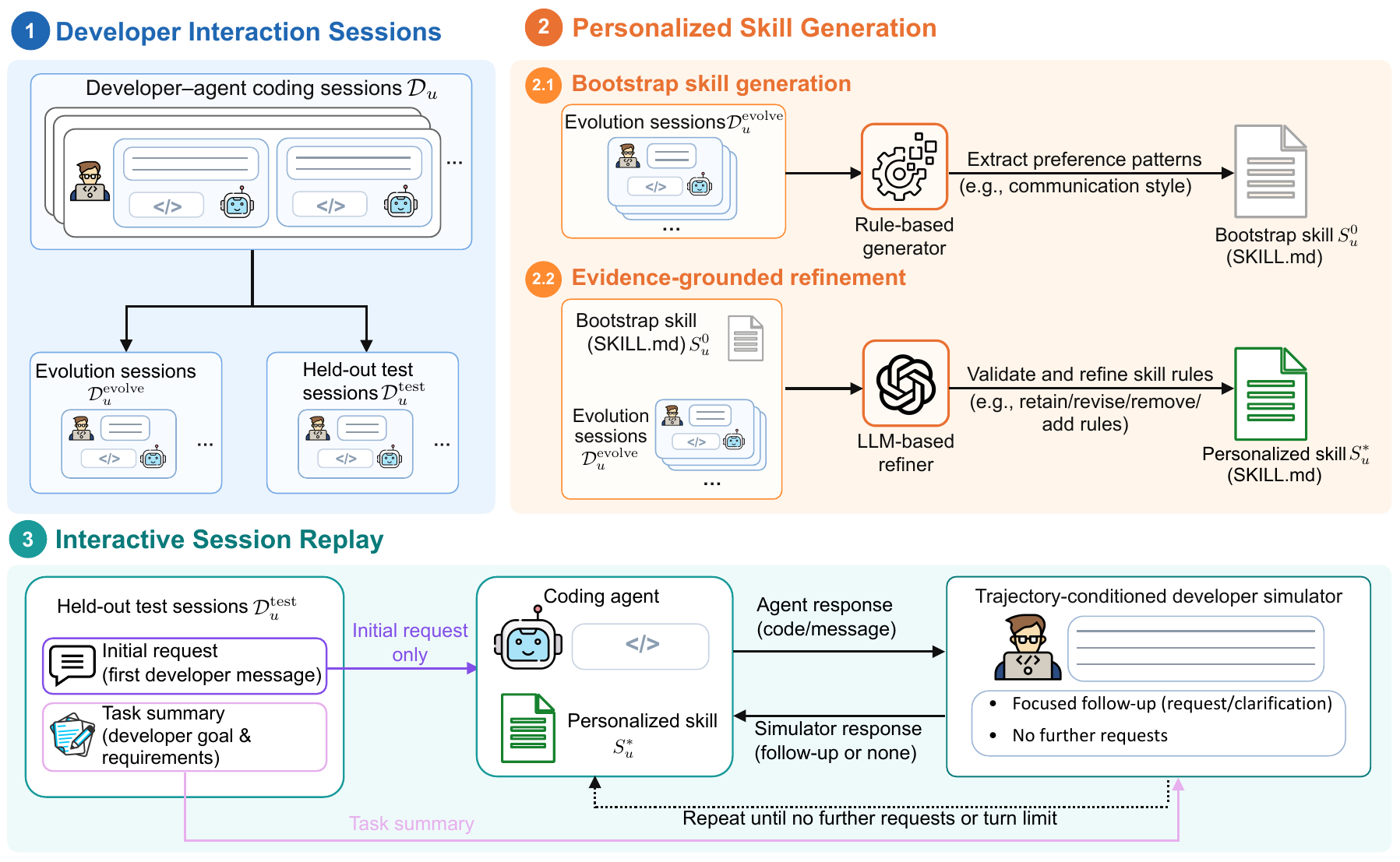}
    \vspace{-.5cm}
    \caption{Overview of the personalized skill generation and interactive session replay framework.}
    \label{fig:overview}
\end{figure*}

\paragraph{Contributions} We propose a framework for generating personalized skills from individual developers' interaction \emph{traces} with coding agents in sessions. For skill construction, we adopt a two-stage procedure: First, we construct a bootstrap skill using predefined rules that organize common preference dimensions into an initial structured skill. Second, we use an ``evolution-set'' of traces to refine the skill based on the developer's interactions with the coding agent, distilling transferable developer-specific preferences by adding, revising, or removing instructions according to the available evidence. At inference time, the generated skills are incorporated into the agent's prompt to guide its behavior without modifying any LLM parameters. For reproducible replay, we summarize the developer's requests from each target session into a concise task summary. Conditioned on the summary, an LLM-based developer simulator progressively reveals requirements and provides feedback after each coding-agent response. We conduct experiments on a subset of SWE-Chat \cite{baumann2026swechat}, using 206 real-world developer–agent interaction sessions from 13 developers, with two skill generators and four coding-agent executors. Results show that the effectiveness of personalized skills varies substantially across generator–executor pairings, with limited evidence of improvement in task success or reductions in developer follow-ups. Further comparison across different skill variants reveals that generic skills distilled across developers achieve the highest average task-completion score and significantly outperform the no-skill baseline, whereas personalized skills are only beneficial when historical interactions contain multiple sessions relevant to the target task. 
%do not yield statistically significant improvements. Additional analysis suggests that personalization is more beneficial  

We state up front that our exploratory work is limited by the amount of data: most publicly available human-AI collaborative coding datasets only contain session traces but not required task files. However, our work requires sessions that can be replayed locally and verifiably, which significantly reduces the amount of available data.
% Overall, our findings suggest that broadly transferable procedural knowledge may offer more reliable benefits than developer-specific preferences when individual interaction histories are limited.

\section{Methodology}

% Figure~\ref{fig:overview} illustrates the overall framework. Given a developer's collection of collaborative coding sessions with the coding agent, we first split them into an evolution set and a held-out test set. We use the evolution sessions to construct a personalized skill by first generating a bootstrap skill and then refining it with evidence extracted from the evolution-set traces. We then replay the held-out test sessions with a trajectory-conditioned developer simulator to evaluate personalized skills across different skill generator–executor pairings and compare coding-agent performance under different skill conditions. 

Figure~\ref{fig:overview} illustrates the overall framework. We partition each developer's coding-agent sessions into an evolution set and a held-out test set. We first construct personalized skills using rule-based bootstrapping and evidence-grounded refinement, then assess their effectiveness via interactive session replay with a trajectory-conditioned developer simulator across different skill conditions and generator–executor pairings.

% \ml{things are somewhat repeated below - can shorten here if we need space}

\subsection{Problem Formulation}

Following prior work, we define a skill $S$ as a human-readable \texttt{SKILL.md} document that encodes reusable, natural-language guidance for a coding agent at inference time, without modifying its underlying model parameters~\citep{ni2026trace2skill,yang2026skillopt}. In this work, a personalized skill captures the recurring preferences and expectations of a particular developer rather than task-specific experience. Formally, let $\mathcal{U}$ denote a set of developers. Each developer $u \in \mathcal{U}$ is associated with a set of coding-agent sessions $\mathcal{D}_u = \{d_{u,1}, \ldots, d_{u,n_u}\}$, where $n_u$ denotes the number of sessions associated with developer $u$. We split $\mathcal{D}_u$ into an evolution set $\mathcal{D}_u^{\mathrm{evolve}}$ and a disjoint, held-out test set $\mathcal{D}_u^{\mathrm{test}}$. We use $\mathcal{D}_u^{\mathrm{evolve}}$ to construct the personalized skill and reserve $\mathcal{D}_u^{\mathrm{test}}$ for generalization evaluation. Specifically, given the evolution sessions of developer $u$, we construct a personalized skill $S_u^{*} = \mathcal{G}(\mathcal{D}_u^{\mathrm{evolve}})$, where $\mathcal{G}$ denotes the personalized skill-generation procedure. 

We evaluate personalized skills on task success when the coding agent has access to these skills on tasks in the held-out test set.

\subsection{Personalized Skill Generation}

Our process for personalized skill generation consists of two stages. We first generate a task-independent bootstrap skill from the evolution sessions, which consists of a series of instructions, and then refine it using evidence from the original trajectories:
\[
S_u^{0}
=
\mathcal{B}\left(\mathcal{D}_u^{\mathrm{evolve}}\right),
\qquad
S_u^{*}
=
\mathcal{R}\left(
S_u^{0},
\mathcal{D}_u^{\mathrm{evolve}}
\right).
\]
Here, \(\mathcal{B}\) denotes bootstrap construction and \(\mathcal{R}\) denotes evidence-grounded refinement, including adding, revising, or removing instructions, which we detail below.

\paragraph{Bootstrap skill generation.}
To provide a stable initialization point and reduce skill content drift during subsequent refinement, we first construct a task-independent bootstrap skill $S_u^{0}$ from the developer's evolution sessions $\mathcal{D}_u^{\mathrm{evolve}}$ using a deterministic, rule-based generator. The generator extracts all developer turns in $\mathcal{D}_u^{\mathrm{evolve}}$ and applies lightweight pattern-matching rules to identify communication style, work preferences, follow-up and correction behavior, and, when explicitly indicated, validation and commit preferences. The generator also selects a small set of representative developer turns to illustrate interaction style. These elements are assembled into a structured \texttt{SKILL.md} file containing sections for scope, communication, work style, follow-up handling, validation, and representative interaction examples.

\paragraph{Evidence-grounded refinement.}

To improve the reliability of the bootstrap skill and reduce overgeneralization, we next provide the bootstrap skill and the corresponding evolution sessions to an LLM-based refiner. Inspired by prior approaches that distill reusable behavioral guidance from interaction traces and iteratively refine persistent instructions based on observed evidence~\citep{ni2026trace2skill,yang2026skillopt,wang2026skillgrad}, we adopt an evidence-grounded refinement process to improve the bootstrap skill. Unlike the bootstrap stage, which prioritizes coverage, the refinement stage prioritizes evidence-based validation and generalizability. The refiner treats the evolution sessions $\mathcal{D}_u^{\mathrm{evolve}}$ as the primary source of evidence and the bootstrap skill as an initial set of candidate rules. It verifies each candidate rule against the evolution-set traces, retaining, revising, or removing existing rules, meanwhile adding recurring preferences missed during bootstrap generation. To reduce overfitting, the refiner retains a rule only when it is supported by at least two independent developer turns from different evolution sessions. The refinement stage focuses primarily on communication style, work style, and follow-up handling while avoiding assumptions about programming languages, frameworks, interfaces, or task types. The resulting skill guides the agent's behavior without overriding the active developer request or the current task details. Accordingly, it must not introduce new task requirements, prescribe environment-specific commands, or justify implementing less than the developer requested. The final output is a compact, task-independent \texttt{SKILL.md} file intended to generalize to unseen sessions from the same developer. 

% \ml{this process is quite similar to those agent harness papers, right? should cite and say inspired by xxx}

\subsection{Interactive Session Replay}

Faithful evaluation of personalized skills requires the coding agent to use them on a developer's tasks. However, a recorded developer-agent session cannot be replayed verbatim, since the agent, armed with new skills, may follow a different trajectory from the agent in the original interaction. Therefore, using the original developer follow-up messages at each turn does not lead to a faithful representation of human behavior. Following prior work~\citep{wu2026swetogether}, we replay each held-out test session with an LLM-based developer simulator. The simulator draws on the original developer's task specifications but decides when and how to follow up based on the current session's trajectory. We detail components of our replay setup below. 

\paragraph{Task summary construction.}

For each held-out test session $d \in \mathcal{D}_u^{\mathrm{test}}$, we generate a task summary from all developer-authored messages in the original session. The summary captures the developer's general goal and the requirements explicitly stated during the interaction. It gives the developer simulator a complete account of the developer's stated requirements while enabling the evaluated agent to discover later requirements through subsequent interaction. Since the summary is extracted from developer messages only, it excludes the original agent's reasoning, actions, and implementation choices. Only the developer simulator receives the task summary, whereas the coding agent receives the original, ground-truth first-turn developer request and encounters later requirements through interactive session replay.

\paragraph{Trajectory-conditioned developer simulation.}

After each agent response, the developer simulator receives the task summary and the current conversation with the agent. The simulator reviews the agent's latest response and determines whether it has addressed the developer's explicit requests. It then takes one of two actions: issue a focused follow-up for the most important unresolved requirement or return \texttt{No further requests.} For each follow-up, the simulator refers to what the evaluated agent has said or done and focuses on any remaining gaps. This setup keeps the interaction faithful to the current trajectory without changing the scope of the recorded session. 
% We also analyze the effectiveness of our developer simulation approach in Appendix \ref{app:developer_simulator}.
%. It does not copy the original follow-up messages or preserve their original timing

\subsection{Skill-Conditioned Replay}

For each held-out test session $d \in \mathcal{D}_u^{\mathrm{test}}$, we replay the reconstructed interaction under each skill condition. The coding agent receives the corresponding skill before the developer's original first-turn request. The skill is kept unchanged throughout the entire interactive replay session.

\section{Experimental Setup}

In this section, we conduct experiments to evaluate the effectiveness of our personalized skill generation framework using the proposed replay evaluation setup. 

\subsection{Dataset}

\begin{figure}[!tpbh]
    \centering
    \includegraphics[width=1\columnwidth]{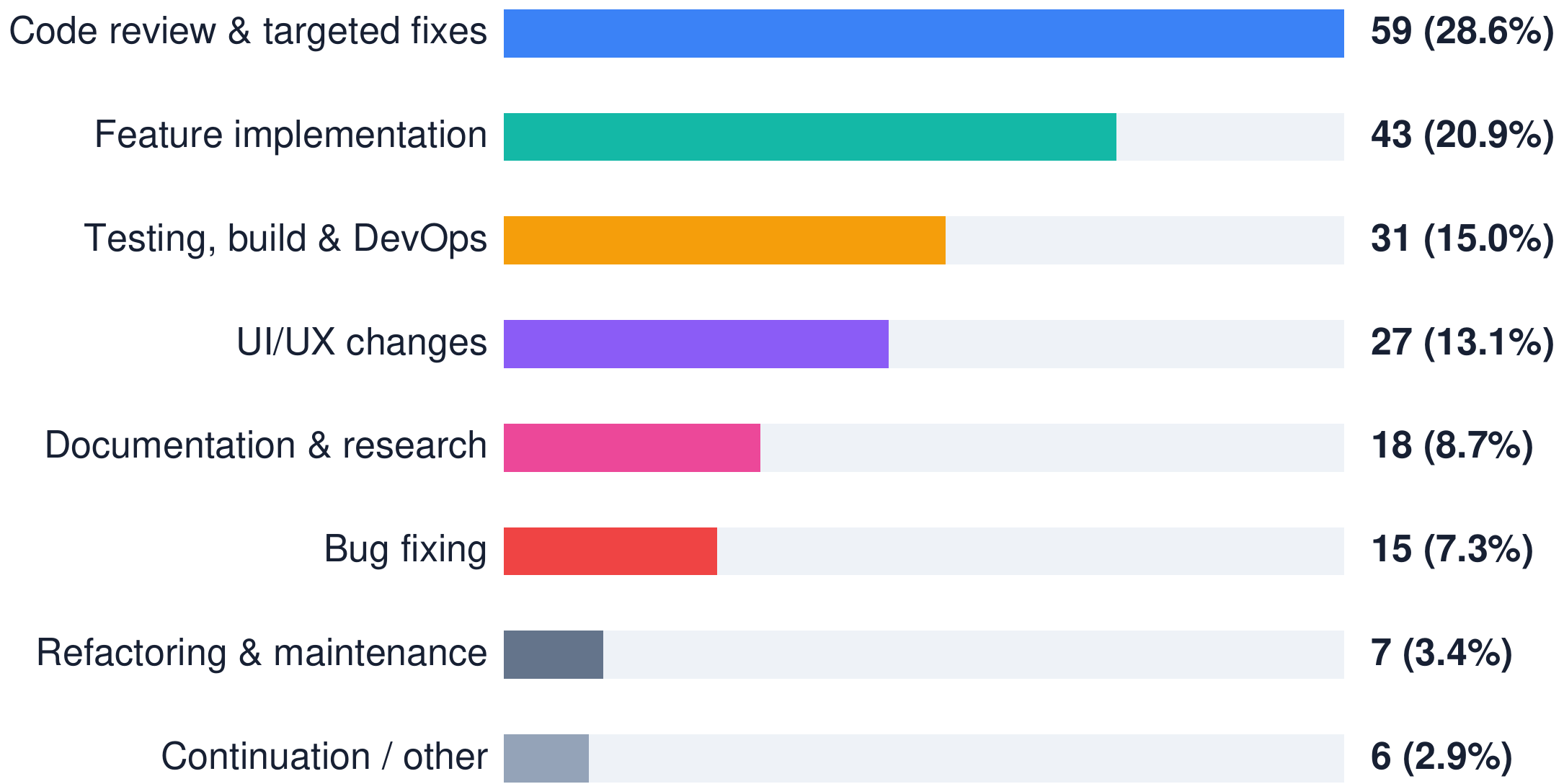}
    \caption{Distribution of task categories in our replayable SWE-chat subset.}
    \label{fig:task_distribution}
    \vspace{-0.5cm}
\end{figure}

Since this work focuses on developer-agent interaction, we evaluate the effectiveness of personalized skills on SWE-chat~\citep{baumann2026swechat}, which contains 8,866 public command line (CLI) interface coding-agent sessions collected through Entire.io from public GitHub repositories between January and June 2026. To enable fully reproducible and verifiable replay, we remove sessions with incomplete interactions, inaccessible or unrecoverable repository states, private dependencies, missing files, or no substantive code edits. For each remaining session, we reconstruct the codebase as it existed before the agent began the task by checking out the parent of the commit containing the agent's final code changes, and create an isolated worktree for execution and validation. We also exclude answer-only, read-only, empty-tool and review-only sessions, and retain only developers with at least three valid tasks. After this filtering, the dataset contains \emph{only 206 sessions from 13 developers}. 
%We note that this strict filtering criterion establishes a rigorous evaluation setup but limits our ability to draw definitive conclusions. 
For each developer, we randomly split the sessions into an 80\% evolution set and a 20\% held-out test set to evaluate generalization to unseen tasks from the same developer, resulting in 164 evolution sessions and 42 test sessions. Because this split is random rather than chronological, our evaluation measures cross-task generalization to held-out sessions, assuming that developer preferences are static and do not change over time. The distribution of task categories in the dataset is shown in Figure~\ref{fig:task_distribution}. This set covers a diverse range of SWE tasks, such as code review and targeted fixes (28.6\%), feature implementation (20.9\%) and testing/build/DevOps (15.0\%).

\begin{table*}[!tbph]
\centering
\small
\setlength{\tabcolsep}{3pt}

\begin{tabularx}{\textwidth}{
    l
    >{\raggedright\arraybackslash}X
    c
    >{\centering\arraybackslash}X
    c
}
\toprule
\cellcolor{HeaderBlue}\textbf{Coding-agent Executor}
& \cellcolor{HeaderBlue}\textbf{Condition}
& \cellcolor{HeaderBlue}\textbf{Score} $\uparrow$
& \cellcolor{HeaderBlue}\textbf{Follow-up Rate} $\downarrow$
& \cellcolor{HeaderBlue}\textbf{Win/Tie/Loss} \\
\midrule

\multicolumn{5}{c}{
\textit{Skill Generator: GPT-5.6 Sol-low}
} \\
\midrule

% ========== Sol-low ==========
\rowcolor{SolLowBg}
& (A) No skill
& $65.27_{\pm 1.94}$
& $24.76\%$ $(52/210)$
& -- \\

\rowcolor{SolLowBg}
\multirow{-2}{*}{GPT-5.6 Sol-low}
& (B) Personalized skill
& $67.64_{\pm 2.71}$
& $30.95\%$ $(65/210)$
& $51.90 / 7.62 / 40.48\%$ $(109/16/85)$ \\

\addlinespace[4pt]

% ========== Sol-high ==========
\rowcolor{SolHighBg}
& (A) No skill
& $68.59_{\pm 4.39}$
& $27.62\%$ $(58/210)$
& -- \\

\rowcolor{SolHighBg}
\multirow{-2}{*}{GPT-5.6 Sol-high}
& (B) Personalized skill
& $66.71_{\pm 7.34}$
& $25.71\%$ $(54/210)$
& $47.62 / 7.62 / 44.76\%$ $(100/16/94)$ \\

\addlinespace[4pt]

% ========== Sol-max ==========
\rowcolor{SolMaxBg}
& (A) No skill
& $70.22_{\pm 1.33}$
& $26.19\%$ $(55/210)$
& -- \\

\rowcolor{SolMaxBg}
\multirow{-2}{*}{GPT-5.6 Sol-max}
& (B) Personalized skill
& $72.04_{\pm 2.18}$\textsuperscript{*}
& $29.05\%$ $(61/210)$
& $53.33/ 11.43 / 35.24 \%$ $(112/24/74)$ \\

\addlinespace[4pt]

% ========== Astra-med ==========
\rowcolor{AstraMedBg}
& (A) No skill
& $65.40_{\pm 3.35}$
& $32.38\%$ $(68/210)$
& -- \\

\rowcolor{AstraMedBg}
\multirow{-2}{*}{GPT-6 Astra-medium}
& (B) Personalized skill
& $65.59_{\pm 4.34}$
& $37.14\%$ $(78/210)$
& $47.62 / 12.86/ 39.52 \%$ $(100/27/83)$ \\

\midrule

\multicolumn{5}{c}{
\textit{Skill Generator: GPT-6 Astra-medium}
} \\
\midrule

% ========== Sol-low ==========
\rowcolor{SolLowBg}
& (A) No skill
& $63.26_{\pm 1.15}$
& $13.81\%$ $(29/210)$
& -- \\

\rowcolor{SolLowBg}
\multirow{-2}{*}{GPT-5.6 Sol-low}
& (B) Personalized skill
& $63.07_{\pm 0.62}$
& $12.38\%$ $(26/210)$
& $48.10 / 9.52 / 42.38\%$ $(101/20/89)$ \\

\addlinespace[4pt]

% ========== Sol-high ==========
\rowcolor{SolHighBg}
& (A) No skill
& $63.14_{\pm 4.08}$
& $10.95\%$ $(23/210)$
& -- \\

\rowcolor{SolHighBg}
\multirow{-2}{*}{GPT-5.6 Sol-high}
& (B) Personalized skill
& $66.41_{\pm 5.90}$
& $11.43\%$ $(24/210)$
& $55.24 / 10.95 / 33.81\%$ $(116/23/71)$ \\

\addlinespace[4pt]

% ========== Sol-max ==========
\rowcolor{SolMaxBg}
& (A) No skill
& $71.43_{\pm 2.82}$
& $30.48\%$ $(64/210)$
& -- \\

\rowcolor{SolMaxBg}
\multirow{-2}{*}{GPT-5.6 Sol-max}
& (B) Personalized skill
& $72.33_{\pm 3.62}$
& $25.71\%$ $(54/210)$
& $47.14 / 7.14 / 45.71\%$ $(99/15/96)$ \\

\addlinespace[4pt]

% ========== Astra-med ==========
\rowcolor{AstraMedBg}
& (A) No skill
& $64.95_{\pm 2.94}$
& $35.71\%$ $(75/210)$
& -- \\

\rowcolor{AstraMedBg}
\multirow{-2}{*}{GPT-6 Astra-medium}
& (B) Personalized skill
& $65.91_{\pm 3.47}$
& $28.10\%$ $(59/210)$
& $52.38 / 11.90 / 35.71\%$ $(110/25/75)$ \\

\bottomrule
\end{tabularx}

\caption{Coding agent task success performance comparison across different agent models and skill generators. We report mean and standard deviation of the task success score. Follow-up Rate denotes the percentage of replay instances in which the simulated developer issues at least one non-terminal follow-up after the initial request, with multiple follow-ups in the same replay counted once. Win/Tie/Loss denotes the percentage of paired replay instances in which the personalized-skill condition scores higher than, equal to, or lower than the no-skill baseline, respectively. \textsuperscript{*}$p<.05$ in a two-sided paired $t$-test over the five seed means against the no-skill baseline.
}
\label{tab:cross_model_results}
\end{table*}

\subsection{Experimental Setup}

We implement skill generation and interactive replay with Codex CLI\footnote{\url{https://github.com/openai/codex}}. We evaluate personalized skills across two skill generators (GPT-5.6 Sol-low and GPT-6 Astra-medium) and four coding-agent executors (GPT-5.6 Sol-low, Sol-high, Sol-max, and GPT-6 Astra-medium). We further compare four skill conditions using GPT-5.6 Sol-low for both skill generation and execution. To ensure a fair comparison, we keep the simulator, scorer, prompts, and replay settings fixed within each configuration. Each held-out task starts from the same repository commit in an isolated worktree and is limited to six developer–agent turns. Appendix~\ref{app:prompts} provides all prompts.

% , while the remaining seven generator–executor configurations compare personalized skills against the no-skill baseline

We consider four conditions: (A) no skill, (B) the target developer's personalized skill, (C) a personalized skill from another random developer and (D) a generic skill pooled across all developers.
% , (E) the target developer's rule-based bootstrap skill before LLM refinement, and (F) a leave-one-developer-out generic skill pooled without the target developer's evolution sessions.
%For Condition~C, we randomly select a personalized skill generated from a different developer. 
For Condition~D, the generic skill is generated from the pooled evolution sessions of all developers, including those of the target developer. 
% In contrast, Condition~F excludes the target developer's evolution sessions to examine whether generic skills remain effective without developer-specific interaction history. 
All other components including the repository state, initial request, task summary, developer simulator, and agent configuration, remain the same across all experimental conditions. After each replay, we collect the interaction trajectory, code changes, and validation evidence, and assess task completion using Claude Opus 5.5 Max\footnote{\url{https://www.anthropic.com/claude-opus-5-5}} as an LLM judge, following the SWE-chat 100-point scoring rubric~\citep{baumann2026swechat}. We repeat the experiment using five random seeds for the within-developer data split to reduce sensitivity to data partitions and improve the statistical robustness.

\subsection{Quantitative Results}
\label{sec:main_results}

\paragraph{Personalized-skill effectiveness depends on the generator--executor pairing.}
Table~\ref{tab:cross_model_results} compares personalized skills across two skill generators and four coding-agent executors. Personalized skills outperform the no-skill baseline in six of the eight configurations, but the improvements vary substantially across model pairings. For example, with the GPT-5.6 Sol-high executor, skills generated by GPT-5.6 Sol-low yield a decrease in average score ($-1.88$), whereas skills generated by GPT-6 Astra-medium yield an improvement ($+3.27$). Only one of the eight comparisons reaches statistical significance (GPT-5.6 Sol-max with GPT-5.6 Sol-low-generated skills, $p=.031$). These findings suggest that personalized-skill effectiveness depends heavily on the generator--executor pairing rather than consistently improving coding-agent performance.

\begin{table*}[!tbph]
\centering
\small
\setlength{\tabcolsep}{3pt}

\begin{tabularx}{\textwidth}{
    l
    c
    >{\centering\arraybackslash}X
    c
}
\toprule
\cellcolor{HeaderBlue}\textbf{Condition}
& \cellcolor{HeaderBlue}\textbf{Score} $\uparrow$
% & \cellcolor{HeaderBlue}\textbf{$\Delta$ vs.\ A} $\uparrow$
& \cellcolor{HeaderBlue}\textbf{Follow-up Rate} $\downarrow$
& \cellcolor{HeaderBlue}\textbf{Win/Tie/Loss}\\
\midrule

(A) No skill
    & $65.27_{\pm 1.94}$
    % & --
    & $24.76\%$ $(52/210)$
    & -- \\

\rowcolor{AlternateGray}
(B) Personalized skill
    & $67.64_{\pm 2.71}$
    & $30.95\%$ $(65/210)$ 
    & $51.90 / 7.62 / 40.48\%$ $(109/16/85)$ \\

(C) Random developer skill
    & $67.75_{\pm 3.68}$
    & $27.62\%$ $(58/210)$
    & $50.95 / 10.48 / 38.57\%$ $(107/22/81)$ \\

\rowcolor{BestGreen}
(D) Generic skill
    & $\mathbf{68.38_{\pm 1.72}}$\textsuperscript{*}
    & $30.00\%$ $(63/210)$
    & $\mathbf{55.71 / 7.62 / 36.67\%}$ $(117/16/77)$ \\

% (E) Rule-based bootstrap
%     & $67.49_{\pm 3.00}$\textsuperscript{*}
%     & $30.95\%$ $(65/210)$
%     & $53.81 / 8.10 / 38.10\%$ $(113/17/80)$ \\

% (F) Leave-one-out generic skill
%     & $68.37_{\pm 1.88}$\textsuperscript{*}
%     & $30.00\%$ $(63/210)$
%     & $57.62 / 8.57 / 33.81\%$ $(121/18/71)$ \\

\bottomrule
\end{tabularx}
% \caption{Task-completion performance under six skill conditions. Follow-up Rate denotes the percentage of replay instances in which the simulated developer issues at least one non-terminal follow-up after the initial request, with multiple follow-ups in the same replay counted once. Win/Tie/Loss denotes the percentage of paired replay instances in which a condition scores higher than, equal to, or lower than the no-skill baseline, respectively. Best results are shown in \textbf{bold}.}
\caption{Task success comparison across all skill conditions.
(A) uses no skill; 
(B) uses a developer's personalized skill;
(C) uses the personalized skill of another random developer;
(D) uses one generic skill pooled from all developers' training sessions;
% (E) uses the target developer's rule-based bootstrap skill without LLM refinement;
% (F) uses a generic skill pooled from all other developers, excluding every training session of the target developer.
% Score is the mean task-completion score over five seeds ($\pm$ standard deviation across seeds), computed on 210 paired replay instances.
% Follow-up Rate denotes the percentage of replay instances in which the simulated developer issues at least one non-terminal follow-up after the initial request, with multiple follow-ups in the same replay counted once.
% Win/Tie/Loss denotes the percentage of paired replay instances in which a condition scores higher than, equal to, or lower than the no-skill baseline, respectively.
Best results are shown in \textbf{bold}. \textsuperscript{*}$p<.05$ in a two-sided paired $t$-test over the five seed means against Condition~A.}
\label{tab:main_results}
\end{table*}

\paragraph{Personalized-skill effectiveness depends on the generator--executor pairing.}
Table~\ref{tab:cross_model_results} compares personalized skills across two skill generators and four coding-agent executors. Personalized skills achieve higher average task-completion scores than the no-skill baseline in six of the eight configurations, but their effectiveness varies substantially across model pairings. For example, with the GPT-5.6 Sol-high executor, skills generated by GPT-5.6 Sol-low yield a decrease in average score (-1.88), whereas those generated by GPT-6 Astra-medium achieve an improvement (+3.27). However, only one of the eight comparisons reaches statistical significance (GPT-5.6 Sol-max with GPT-5.6 Sol-low-generated skills, $p=.031$). These findings suggest that personalized-skill effectiveness depends heavily on the generator--executor pairing rather than consistently improving coding-agent performance.

\paragraph{Generic skills yield the strongest and most reliable gains.}
Table~\ref{tab:main_results} compares the no-skill baseline with three skill conditions under the primary generator--executor configuration. The generic skill achieves the highest task-completion score (68.38), the largest improvement over the no-skill baseline (+3.11), and a win rate of 55.71\%. This improvement is statistically significant (paired $t$-test, $p=.003$), whereas the personalized skill yields a smaller, non-significant gain (+2.37, $p=.084$). 
% Moreover, the leave-one-developer-out generic skill achieves nearly identical performance (68.37, $p=.009$), with no significant difference from the generic skill ($p=.990$), suggesting that its effectiveness does not depend on including the target developer's interaction history.

In contrast, a skill generated for another developer (67.75) performs similarly to the target developer's personalized skill (67.64), providing limited evidence of a developer-specific advantage. Furthermore, neither the
other-developer skill ($p=.838$) nor the generic skill
($p=.392$) differs significantly from the personalized
skill in direct paired $t$-tests. And the generic skill increases the follow-up rate from 24.76\% to 30.00\%, suggesting that its benefits lie primarily in improving task completion rather than reducing developer pushback.
% The rule-based bootstrap skill also achieves a comparable score (67.49), significantly outperforming the no-skill baseline ($p=.035$). However, LLM-based refinement yields only marginal improvement over the bootstrap (+0.15), with no statistically significant difference ($p=.886$). Both variants have the same follow-up rate (30.95\%), suggesting that refinement does not reduce developer follow-ups.

\section{Detailed Analysis}

We further examine the impact of relevant evolution sessions,
evaluate the grounding of personalized skills through human
evaluation, and assess the developer simulator. We also conduct
ablation studies on skill generation to assess their individual
contributions. Additional analyses of interaction behavior,
skill content, and developer-level variation are provided in
Appendices~\ref{app:Interaction_and_Execution_Behavior},
\ref{app:Skill_Content_Analysis}, and
\ref{app:developer-Level_var}, respectively.

\subsection{Impact of Relevant Evolution Sessions}
\label{app:relevance_analysis}

\begin{table}[!tbph]
\centering
\small
\setlength{\tabcolsep}{4pt}  % 控制列间距

\begin{tabular}{lrrrr}
\toprule
\textbf{Relevant Sessions} & \textbf{Num.} & $\mathbf{B-A}$ & $\mathbf{B-C}$ & $\mathbf{B-D}$ \\
\midrule
0        & 3  & $-6.33$  & $+15.00$ & $-8.00$ \\
1--2     & 16 & $0.00$   & $-1.38$  & $-3.81$ \\
3--5     & 11 & $+0.10$  & $+0.20$  & $-7.20$ \\
$\geq 6$ & 12 & $+10.17$ & $+8.92$  & $+5.67$ \\
\bottomrule
\end{tabular}
\caption{Performance of relevant evolution sessions. A: no skill; B: personalized skill; C: random developer skill; D: generic skill.}
\label{tab:history_overlap}
\end{table}

% We examine whether the number of sessions used for skill creation in the evolution set impacts the effectiveness of personalized skills. For each of the 42 held-out tasks in one random seed, we review all evolution sessions from the same developer and identify those that are semantically related to the held-out task. We then group the held-out tasks into four bins (0, 1-2, 3-5, and $\geq 6$ relevant sessions) to analyze how the effectiveness of personalized skills varies with the number of relevant evolution sessions. As reported in Table~\ref{tab:history_overlap}, the effectiveness of the personalized skill varies with the amount of relevant evolution sessions. When at least six relevant sessions are available, the personalized skill substantially outperforms both the no-skill baseline ($B-A=+10.17$) and the random developer skill ($B-C=+8.92$), and even surpasses the generic skill ($B-D=5.67$). In contrast, when fewer than six relevant sessions are available, the personalized skill shows little or no advantage over either the no-skill baseline or the random developer skill, while the generic skill performs better in all three groups. This result suggests that personalized skills become effective only when a developer's evolution set contains a sufficient number of examples that are relevant to held-out evaluation tasks. Therefore, it is possible that as larger-scale, replayable and verifiable developer-agent interaction data becomes available, the potential of personalized skills can be fully realized. 

We examine whether the number of relevant evolution sessions affects personalized skill effectiveness. For each of the 42 held-out tasks in one random seed, we identify semantically related evolution sessions from the same developer and group tasks into four bins (0, 1--2, 3--5, and $\geq 6$ relevant sessions). As reported in Table~\ref{tab:history_overlap}, the effectiveness of the personalized skill varies with the amount of relevant evolution sessions. When at least six relevant sessions are available, the personalized skill substantially outperforms both the no-skill baseline ($B-A=+10.17$) and the random developer skill ($B-C=+8.92$), and even surpasses the generic skill ($B-D=5.67$). In contrast, when fewer than six relevant sessions are available, the personalized skill shows little or no advantage over either the no-skill baseline or the random developer skill, while the generic skill performs better in all three groups. This result suggests that personalized skills become effective only when a developer's evolution set contains a sufficient number of examples that are relevant to held-out evaluation tasks. Therefore, it is possible that as larger-scale, replayable and verifiable developer-agent interaction data becomes available, the potential of personalized skills can be fully realized.

\subsection{Human Evaluation}
\label{app: Human_Evaluation}

\begin{table}[!tbph]
\centering
\footnotesize
\setlength{\tabcolsep}{4pt}
\begin{tabular}{lccc}
\toprule
\textbf{Label}
& \textbf{Score}
& \textbf{Ann.~1}
& \textbf{Ann.~2} \\
\midrule
\textit{Contradicted}
& 1 & 0 (0.0\%) & 1 (1.1\%) \\
\textit{Insufficient evidence}
& 2 & 11 (12.1\%) & 11 (12.1\%) \\
\textit{Weakly supported}
& 3 & 7 (7.7\%) & 22 (24.2\%) \\
\textit{Supported}
& 4 & 73 (80.2\%) & 57 (62.6\%) \\
\midrule
\multicolumn{2}{l}{Pearson's $r$}
& \multicolumn{2}{c}{0.565} \\
\bottomrule
\end{tabular}
\caption{
Human evaluation results of personalized skill claims ($n=91$).
Entries report count (percentage).
}
\label{tab:human_eval}
\end{table}

To assess whether personalized skills faithfully capture developer preferences, we conduct a claim-level human evaluation of 10 personalized skills comprising 95 claims. We recruit two computer science PhD students independently assess each claim against the corresponding developer's evolution-set interactions. We provide further details on skill sampling and annotation procedures in Appendix~\ref{app: Human_Evaluation}. We exclude four non-behavioral or generic claims, leaving 91 claims for analysis. As shown in Table~\ref{tab:human_eval}, the annotators rate 87.9\% and 86.8\% of claims as supported or weakly supported, respectively, with moderate agreement (Pearson's $r=0.565$). These findings suggest that most extracted claims reflect developers' observed behavior, although such alignment does not necessarily improve performance on held-out tasks.

% To assess whether personalized skills faithfully capture developer preferences, we conduct a claim-level human evaluation of 10 personalized skills generated using GPT-6 Astra-medium. We sample two from each of five random seeds, with each pair drawn from different developers. Each skill is decomposed into rule-level claims, yielding 95 claims in total. We recruit two computer science PhD students who regularly use coding agents to independently evaluate each claim against the corresponding developer's anonymized evolution-set interactions, without access to held-out sessions or evaluation results. 

% The skill claims are classified as \textit{supported},
% \textit{weakly supported}, \textit{insufficient evidence},
% \textit{contradicted}, or \textit{not behavioral or generic}. We exclude four claims labeled \textit{not behavioral or generic} by either annotator, as they fall outside the ordinal support scale, leaving 91 claims for analysis. As shown in Table \ref{tab:human_eval}, the two annotators rate 87.9\% and 86.8\% of claims as supported or weakly supported, respectively. Their ordinal ratings exhibit a moderate correlation (Pearson's $r=0.565$), with 96.7\% of ratings differing by at most one support level. These findings suggest that most extracted claims are grounded in developers' interactions, although such support does not necessarily translate into improved performance on held-out tasks.

\subsection{Effectiveness of Developer Simulator}
\label{app:developer_simulator}

% \begingroup
% \setlength{\columnsep}{10pt}
% \setlength{\intextsep}{-1pt}

\begin{figure}
    \centering
    \includegraphics[width=\linewidth]{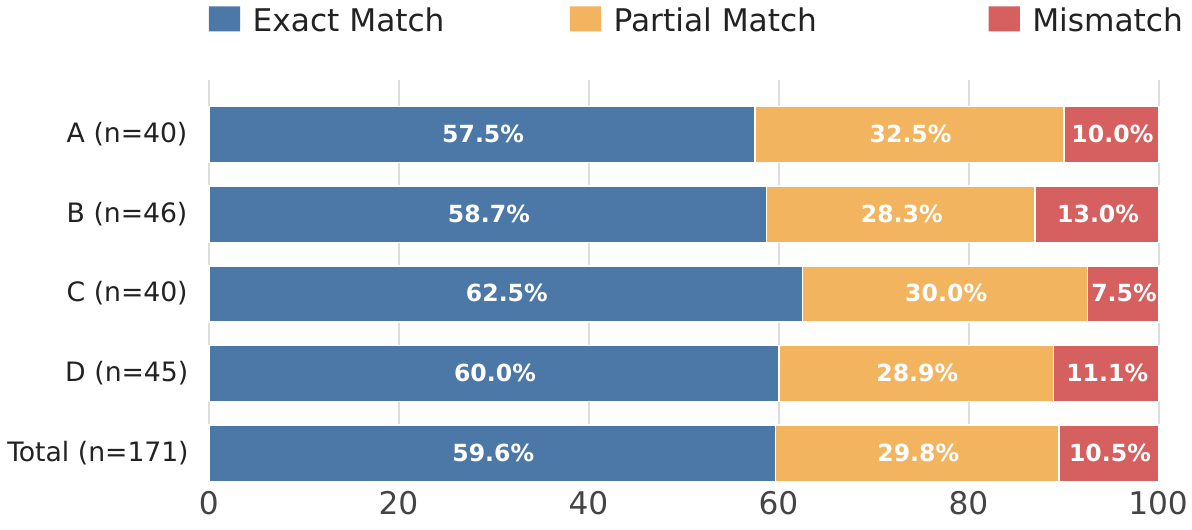}
    \caption{Semantic consistency between real developer messages and
    simulator-generated follow-ups. A: no skill; B: personalized skill; C: random developer skill; D: generic skill.}
    \label{fig:simulator_consistency}
\end{figure}

We also examine the authenticity of the proposed human developer simulator. We compare simulated follow-ups with real developer messages. Each pair of messages is classified by an LLM-as-a-judge evaluator according to a predefined semantic-matching rubric. We define three match levels. An \emph{exact match} indicates that the simulated follow-up has all requirements in the real developer follow-up, allowing minor differences in wording or the consolidation of multiple turns. A \emph{partial match} indicates that the two follow-ups share at least one requirement, but the simulated follow-up omits or introduces other requirements. A \emph{mismatch} indicates that the two follow-ups contain different requirements with no substantive overlap with the human follow-up. Across five seeds, 171 of the 210 replay instances contained substantive follow-ups from both the real developer and the simulator. As shown in Figure~\ref{fig:simulator_consistency}, among these instances, 59.65\% were exact matches, 29.82\% were partial matches, and 10.53\% were mismatches. Thus, 89.47\% were at least partially semantically consistent with real developer messages. This rate ranges from 86.96\% to 92.50\% across the four experimental conditions, suggesting that our replay simulator generally captures the requirements expressed by developers across different skill conditions.

\subsection{Ablation Studies}
\label{app:Ablation Study}

Table~\ref{tab:ablation} examines the effects of LLM-based refinement and in generic skill generation and how including sessions from a developer helps task success for them. LLM-based refinement provides only marginal improvement over rule-based bootstrapping (67.64 vs.\ 67.49, $p=0.886$), suggesting limited additional benefit from evidence-grounded refinement. Similarly, excluding the target developer's evolution sessions from generic skill generation yields nearly identical performance (68.37 vs.\ 68.38, $p=0.990$), suggesting that generic skills benefit primarily from broadly transferable guidance rather than developer-specific information.

\begin{table}[tp]
\centering
\small
\setlength{\tabcolsep}{4pt}
\begin{tabularx}{\columnwidth}{
    >{\raggedright\arraybackslash}X
    c
}
\toprule
\textbf{Condition}
& \textbf{Score} $\uparrow$ \\
\midrule

\multicolumn{2}{l}{
    \textit{Effect of LLM-based refinement}
} \\
(E) Rule-based bootstrap
    & $67.49_{\pm 3.00}$ \\
(B) Personalized skill
    & $67.64_{\pm 2.71}$ \\

\midrule
\multicolumn{2}{l}{
    \textit{Effect of target-developer inclusion}
} \\
(F) Leave-one-out generic
    & $68.37_{\pm 1.88}$ \\
(D) Generic skill
    & $68.38_{\pm 1.72}$ \\

\bottomrule
\end{tabularx}
\caption{
Ablation study of LLM-based refinement and
target-developer inclusion in generic skill generation.
}
\label{tab:ablation}
\end{table}

\section{Conclusions and Future Work}

In this paper, we investigated whether personalized skills distilled from evolution-set developer-agent interactions can improve coding-agent performance on held-out tasks. Experimental results showed limited benefits from developer-specific personalization: personalized skills yielded small and inconsistent gains, while a generic skill pooled across developers provided the largest and most consistent improvement. Further analyses suggested that personalization is more effective when human developers have repeated tasks and their preferences are manifested often enough. We also used a human evaluation to verify the faithfulness of the skill distillation process. 
%limited per-developer evolution sets may make it difficult to identify stable and transferable developer preferences, although personalization appears more promising when multiple relevant evolution-set examples are available.
Overall, our findings suggest that while skills can improve coding agents, benefits arise primarily from reusable guidance shared across developers rather than from personalization. 

As more real-world human-AI interaction datasets become available,
there are promising avenues for future work. First, one can examine whether longer evolution-set traces, with extensive follow-up after the first turn, support more reliable personalization. Second, one can investigate other methods for integrating skills into coding agents instead of including them as part of the prompt. Since tasks we analyzed are GitHub-related work, we do not formulate skills as reusable Python functions, but that should be studied in tasks where they are applicable. Third, one can explore more adaptive mechanisms for skill utilization, such as retrieving task-relevant guidance from a repository of skills mined across developers and tasks. Combining it with a small set of developer-specific preferences may be the most effective setting for a new task.

\section*{Limitations}

Our work has several limitations. First, our study uses a small, highly filtered subset of SWE-chat, comprising 206 replayable sessions from 13 developers. Filtering for repository reconstruction and validation improves experimental control but may bias the sample toward particular developers, repositories, and task types, limiting generalizability. The limited interaction histories per developer also constrain the identification of stable, transferable preferences. Second, our interactive replay framework approximates real developer--agent collaboration. The LLM-based developer simulator may not faithfully reproduce human responses to novel agent trajectories, while task summarization, skill generation, and LLM-based scoring may introduce model-dependent biases. Simulated follow-up rates therefore do not directly measure real-world collaboration efficiency. Additionally, reconstructed repositories may lack historical dependencies or external services, and LLM-based evaluation cannot fully replace executable tests or human assessment. Finally, our findings are limited to natural-language skills injected into agent prompts and may not generalize to other personalization mechanisms, such as retrieval-based memory or online adaptation.

\section*{Ethical Considerations}

Developer interaction traces may contain identifying information, proprietary context, credentials, communication patterns, or other sensitive content, even when associated repositories are public. Distilling personalized skills from these traces may concentrate sensitive information into persistent profiles. Inferred developer preferences may be inaccurate, outdated, or based on insufficient evidence. Applying them automatically risks reinforcing incorrect assumptions, overriding current developer instructions, or creating unequal performance across users with different amounts of interaction history. Therefore, personalized guidance should remain subordinate to explicit user requests and be transparent, inspectable, and correctable by developers. Such profiles should not be used for employee surveillance, performance evaluation, or other consequential decisions. Furthermore, simulated developer feedback and LLM-based scoring may reflect model biases and should not be treated as direct evidence of human satisfaction. Finally, replaying coding sessions involves executing model-generated commands and repository code, requiring isolated, least-privilege environments without production credentials or access to sensitive systems.

\section*{AI Assistance Disclosure}
We used ChatGPT to assist with refining the manuscript's language and clarity. 

\bibliography{custom}

\appendix

\section{Related Work}

% \begin{table}[!tbph]
% \centering
% \small
% \setlength{\tabcolsep}{4pt}
% \renewcommand{\arraystretch}{1.12}

% \begin{tabularx}{\columnwidth}{
%     >{\raggedright\arraybackslash}X
%     c
%     c
% }
% \toprule
% \rowcolor{HeaderBlue}
% \textbf{Support label}
% & \textbf{Annot.\ 1}
% & \textbf{Annot.\ 2} \\
% \midrule

% Supported
%     & 77.9\% & 60.0\% \\
% \rowcolor{AlternateGray}
% Weakly supported
%     & 7.4\% & 23.2\% \\
% Insufficient evidence
%     & 14.7\% & 11.6\% \\
% \rowcolor{AlternateGray}
% Contradicted
%     & 0.0\% & 1.1\% \\
% Not behavioral or generic
%     & 0.0\% & 4.2\% \\

% \midrule
% \rowcolor{SectionGray}
% \textbf{Supported (combined)}
%     & \textbf{85.3\%}
%     & \textbf{83.2\%} \\

% \midrule
% \multicolumn{3}{l}{
%     \textbf{Inter-annotator agreement}
% } \\
% Binary support agreement
%     & \multicolumn{2}{c}{93.7\% / 0.76} \\
% \rowcolor{AlternateGray}
% Five-category agreement
%     & \multicolumn{2}{c}{64.2\% / 0.28} \\

% \bottomrule
% \end{tabularx}

% \caption{Human evaluation of 95 claims from 10 personalized
% skills. Supported (combined) includes supported and weakly
% supported claims. Binary support groups these two labels
% as supported and the remaining labels as unsupported.
% Agreement is reported as percentage agreement / Cohen's $\kappa$.}
% \label{tab:human_eval}
% \end{table}

\subsection{Interactive Coding Agents}

As LLM-based agents have demonstrated increasing capability in solving complex SWE tasks, many studies have developed benchmarks that provide agents with task specifications and evaluate the resulting repository state~\citep{jimenez2024swe,yang2024swe,wang2025openhands}. Most recent work has moved toward more realistic interactive settings \citep{tang2026programming}. For example, SWE-chat collects real coding-agent sessions from open-source development environments, providing an empirical understanding of how AI agents perform in real developer workflows~\citep{baumann2026swechat}. Based on SWE-chat, SWE-Together reconstructs verifiable multi-turn tasks and user feedback from the collected sessions, enabling the evaluation of both final task correctness and the amount of user guidance required during interaction~\citep{wu2026swetogether}. SWE-Interact transforms existing single-turn SWE benchmarks into interactive tasks in which a simulated user progressively reveals requirements and provides feedback during execution, evaluating whether agents can discover user intent, adapt to evolving requirements, and build on their own prior work~\citep{raghavendra2026sweinteract}. Whereas these studies focus on within-session interaction and adaptation, we investigate cross-session adaptation by examining whether experience from a developer's evolution-set sessions can guide agent behavior on held-out tasks.

\subsection{Agent Skills and Skill Evolution}

Agent Skills encode reusable procedural knowledge in structured packages of instructions, code, and supporting resources~\citep{xu2026agent,jiang2026sok,li2026skillsbench,cho2026skillret,zhao2026generative}. Since skills reside outside the model parameters, they provide a lightweight mechanism for adapting agent behavior without parameter updates. Recent work has explored automatically deriving and improving skills from agents' prior execution experience \citep{alzubi2026evoskill,liu2026skillforge,shen2026skillfoundry}. For example, Trace2Skill uses inductive reasoning over agent experience to consolidate multiple execution trajectories into a unified skill directory~\citep{ni2026trace2skill}. SkillOpt iteratively refines natural-language skill documents using rollout feedback and validation-gated edits~\citep{yang2026skillopt}. SkillGrad formulates skill improvement as gradient-descent-like optimization, converting trajectory-level diagnoses into textual gradients that guide structured skill revisions~\citep{wang2026skillgrad}. Some studies further
suggests that skills learned from narrow experience may become overly
specialized, whereas skills distilled from more diverse traces can
transfer more reliably~\citep{belikova2026managing}. These methods primarily aim to extract generalizable knowledge that improves performance across tasks within a domain or benchmark. In contrast, we generate skills from an individual developer's evolution-set interaction traces, focusing on recurring user-specific preferences and expectations rather than broadly applicable task-solving procedures.

\begin{table*}[!tbph]
\centering
\small
\setlength{\tabcolsep}{6pt}
\begin{tabularx}{\textwidth}{
    >{\hsize=2.2\hsize\linewidth=\hsize
      \raggedright\arraybackslash}X
    >{\hsize=0.7\hsize\linewidth=\hsize
      \columncolor{BaselineGray}\centering\arraybackslash}X
    >{\hsize=0.7\hsize\linewidth=\hsize
      \columncolor{PersonalizedBlue}\centering\arraybackslash}X
    >{\hsize=0.7\hsize\linewidth=\hsize
      \columncolor{OtherPurple}\centering\arraybackslash}X
    >{\hsize=0.7\hsize\linewidth=\hsize
      \columncolor{GenericGold}\centering\arraybackslash}X
}
\toprule
\rowcolor{HeaderBlue}
\textbf{Metric}
& \textbf{A}
& \textbf{B}
& \textbf{C}
& \textbf{D} \\
\midrule

\rowcolor{SectionGray}
\multicolumn{5}{l}{\textbf{Interaction and computation}} \\

Agent turns
    & 1.29
    & 1.37
    & 1.33
    & 1.35 \\

Unresolved-requirement follow-ups
    & 0.29
    & 0.37
    & 0.33
    & 0.35 \\

Command/tool calls
    & 8.47
    & 9.46
    & 8.69
    & 8.98 \\

Agent tokens
    & 442{,}096
    & 597{,}120
    & 521{,}219
    & 643{,}578 \\

Execution time (s)
    & 91.76
    & 106.16
    & 99.62
    & 104.57 \\

\midrule
\rowcolor{SectionGray}
\multicolumn{5}{l}{\textbf{Code changes}} \\

Files changed
    & 1.65
    & 2.22
    & 1.79
    & 1.99 \\

Patch churn
    & 29.08
    & 37.11
    & 32.32
    & 37.98 \\

\midrule
\rowcolor{SectionGray}
\multicolumn{5}{l}{\textbf{Testing and validation}} \\

Test command groups
    & 0.56
    & 0.97
    & 0.86
    & 0.93 \\

Validation command groups
    & 1.77
    & 2.23
    & 2.06
    & 2.16 \\

Runs reporting successful validation $\uparrow$
    & 43.1\%
    & \textbf{58.9\%}
    & 54.1\%
    & 57.4\% \\

\bottomrule
\end{tabularx}
\caption{Interaction and execution metrics under the four skill
conditions. Values are averaged across evaluated
replays. A: no skill; B: personalized skill; C: random developer skill;
D: generic skill.}
\label{tab:process_metrics}
\end{table*}

\subsection{Personalized LLM Agents}

Previous work has personalized LLM-based agents using user profiles, persistent memory, and feedback from past interactions~\citep{xu2026personalizedagents,westhausser2025personalized,liang2026pahf,qiu2026autorefine}. These systems typically summarize user information, retrieve relevant preferences during inference, and update stored memories as new feedback become available. AdaMem further studies what information should be retained over long interaction histories~\citep{chen2026adamem}. Personalization has also recently attracted attention in SWE agents. For example, ToM-SWE employs a user-modeling agent to infer developer goals, constraints, and preferences from instructions and interaction history and store them in persistent memory~\citep{zhou2025tom}. Hedwig learns behavioral guidelines from developer interactions across sessions to adjust the autonomy of coding agents~\citep{shukla2026hedwig}. TRACE converts user corrections into persistent, developer-specific rules that can be enforced on future tasks~\citep{zhou2026getting}. AutoSkill identifies recurring preferences and requirements from dialogue histories and encodes them as reusable skills for future interactions~\citep{yang2026autoskill}. Unlike these approaches, our work focuses on learning task-independent personalized skills from developers' evolution-set interaction traces and systematically evaluates whether such skills generalize to held-out tasks from the same developer.

\section{Details of Human Evaluation}
\label{app:Human_Evaluation}
We sample 10 personalized skills generated using GPT-6 Astra-medium, selecting two skills from each of five random seeds, with each pair representing different developers. We decompose these skills into 95 rule-level claims. We recruit two computer science PhD students who regularly use coding agents to independently evaluate each claim using the corresponding developer's anonymized evolution-set interactions. Neither annotator has access to held-out sessions or evaluation results.

We ask annotators to classify each claim as \textit{supported}, \textit{weakly supported}, \textit{insufficient evidence}, \textit{contradicted}, or \textit{not behavioral or generic}. We exclude four claims that either annotator labels as \textit{not behavioral or generic}, since these claims fall outside the ordinal support scale. The remaining 91 claims form the basis of our analysis. The annotators' ratings differ by at most one support level for 96.7\% of these claims.

\section{Interaction and Execution Behavior}
\label{app:Interaction_and_Execution_Behavior}

We further investigate how the skill conditions affect the agent's interaction and execution behavior. As shown in Table~\ref{tab:process_metrics}, a skill generally leads to more extensive execution rather than reducing interaction effort. Compared to the no-skill baseline, all three skill conditions produce more agent turns, follow-ups, generated tokens, tool calls, code changes, and validation attempts. More specifically, personalized skills do marginally increase the task-completion score from 65.02 to 65.99, but at the cost of increasing the average number of tool calls from 8.47 to 9.46 and execution time from 91.76 to 106.16 seconds. It changes more files and produces greater patch churn, while the number of unresolved requirement follow-ups increases from 0.29 to 0.37. Thus, personalized skills do not reduce the amount of interaction or computation required to complete a task. Nevertheless, skills lead to more systematic validation. Conditioned on the personalized skills, the average number of test command groups increases from 0.56 to 0.97 and the proportion of runs reporting successful validation rises from 43.1\% to 58.9\%. The random developer and generic skills show similar trends. The generic skill achieves the highest task-completion score, but also uses the most tokens and produces the greatest patch churn. These results suggest that skills primarily encourage more extensive implementation and validation rather than improving execution efficiency.

\section{Skill Content Analysis}
\label{app:Skill_Content_Analysis}

\begin{table}[!tbph]
\centering
\small
\setlength{\tabcolsep}{4pt}
\renewcommand{\arraystretch}{1.08}

\begin{tabularx}{\columnwidth}{
    >{\raggedright\arraybackslash}X
    >{\columncolor{PersonalizedBlue}\centering\arraybackslash}p{1.9cm}
    >{\columncolor{GenericGold}\centering\arraybackslash}p{1.6cm}
}
\toprule
\rowcolor{HeaderBlue}
\textbf{Skill property}
& \textbf{Personalized}
& \textbf{Generic} \\
\midrule

\rowcolor{SectionGray}
\multicolumn{3}{l}{\textbf{Skill size and length}} \\

Rules per skill
    & 14.15
    & 25.00 \\
Words per rule
    & 16.70
    & 15.32 \\
Rule words per skill
    & 236.38
    & 383.00 \\

\midrule
\rowcolor{SectionGray}
\multicolumn{3}{l}{\textbf{Rule-category distribution}} \\

Communication rules
    & 23.4\%
    & 20.0\% \\
Workflow rules
    & 30.4\%
    & 28.0\% \\
Validation rules
    & 10.3\%
    & 12.0\% \\
Follow-up rules
    & 28.3\%
    & 24.0\% \\
Commit rules
    & 7.6\%
    & 16.0\% \\

\midrule
\rowcolor{SectionGray}
\multicolumn{3}{l}{\textbf{Similarity and specificity}} \\

Similarity to generic skill
    & 0.517
    & 1.000 \\
Similarity between developers
    & 0.443
    & -- \\
Unique developer-specific rules
    & 64.7\%
    & -- \\

\bottomrule
\end{tabularx}
\caption{Content statistics of the personalized and generic skills. The personalized results are averaged across the skills of 13 developers.}
\label{tab:skill_content}
\end{table}

We also compare the 13 personalized skills for each developer and the generic skill across all developers to better understand why the generic skill consistently outperforms personalized skills. As listed in Table \ref{tab:skill_content}, the generic skill contains substantially more rules (25 vs.\ 14.15 on average) and words per skill (383 vs.\ 236.38) than personalized skills. These two types of skills exhibit similar category distributions, with most rules concerning workflow, follow-up handling, and communication. However, the generic skill contains a larger proportion of commit-related rules (16.0\% vs. 7.6\%). Personalized skills are not simply specialized versions of the generic skill; their mean TF-IDF similarity to the generic skill is 0.517, while the mean similarity between different developers' skills is 0.443. Under our rule-based lexical criterion, 64.7\% of personalized rules are unique to a single developer. There is substantial lexical variation in personalized skills across developers. Nevertheless, lexical uniqueness alone does not imply that a rule is broadly applicable or relevant to held-out tasks, nor does it guarantee that the coding agent will follow the rule during execution. These observations suggest that the generic skill benefits from broader coverage of reusable developer practices, whereas many developer-specific rules may have limited opportunities to influence held-out tasks. These observations are consistent with our earlier hypothesis in Section~\ref{sec:main_results} that limited per-developer evolution sets may constrain the coverage of personalized skills, whereas pooling interactions across developers enables the generic skill to capture a broader range of reusable guidance.

\begin{figure*}[!tpbh]
    \centering
    \includegraphics[width=0.9\textwidth]{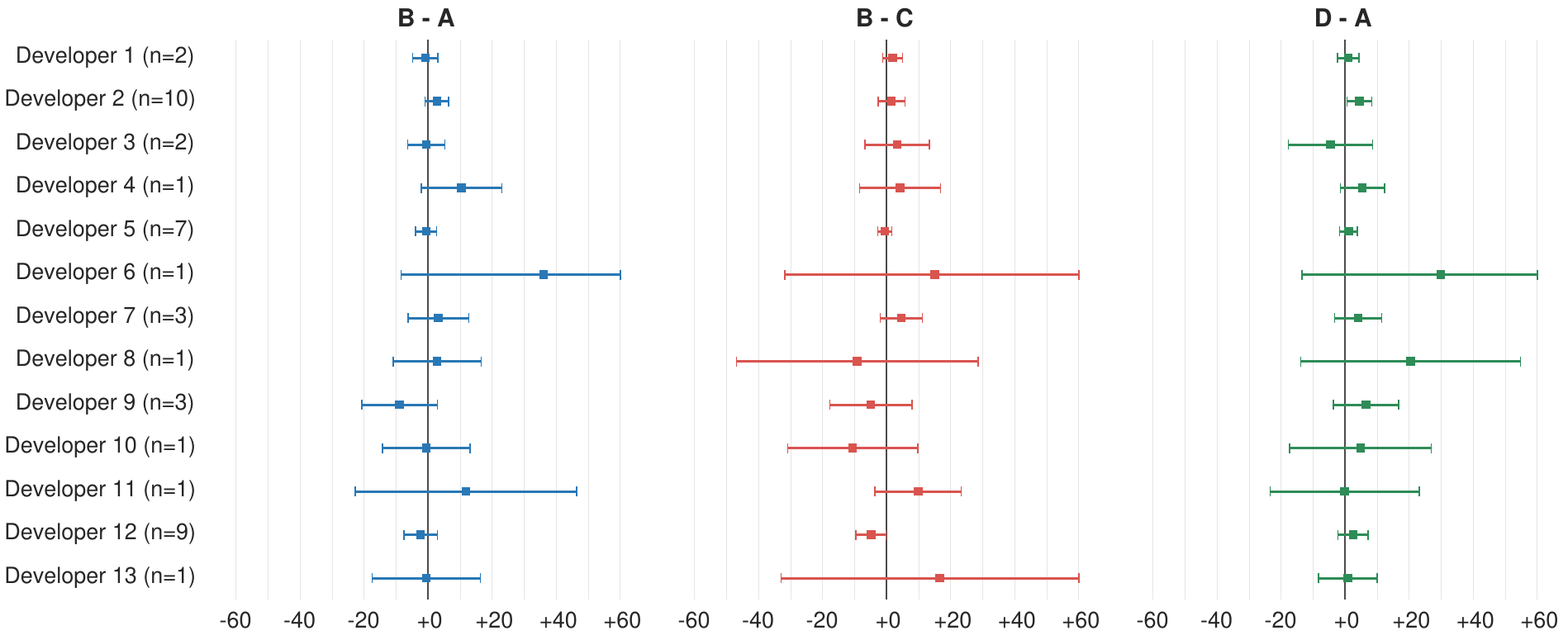}
    \caption{Developer-level differences in task-completion scores across skill conditions (A: no skill; B: personalized skill; C: other-developer skill; D: generic skill). Square markers indicate mean score differences, and horizontal bars show descriptive 95\% intervals.}
    \label{fig:user_level_forest}
\end{figure*}

\section{Developer-Level Variation in Skill Effectiveness}
\label{app:developer-Level_var}

We further examine how skill effectiveness varies across developers. As shown in Figure~\ref{fig:user_level_forest}, the effect of personalized skills varies substantially. Comparing the personalized-skill and no-skill conditions ($B-A$), only 6 of the 13 developers achieve higher task-completion scores under the personalized-skill condition than under the no-skill baseline. Similarly, comparing the target developer's skill with a skill distilled from another developer ($B-C$) reveals no consistent evidence of a personalization benefit. Although the personalized skill achieves higher task-completion scores than the mismatched skill for 8 of the 13 developers, the largest improvements tend to occur for developers with fewer held-out sessions, whose estimates also exhibit substantially wider confidence intervals. In contrast, the generic skill shows more consistent improvements, outperforming the no-skill baseline for 11 of the 13 developers. This observation suggests that limited developer-agent interaction session trace data make it difficult to reliably identify developer-specific preferences that generalize to held-out tasks.

\section{Prompts used in the personalized skill generation framework}
\label{app:prompts}

\begin{figure*}[!tpbh]
    \centering
    \includegraphics[width=\textwidth]{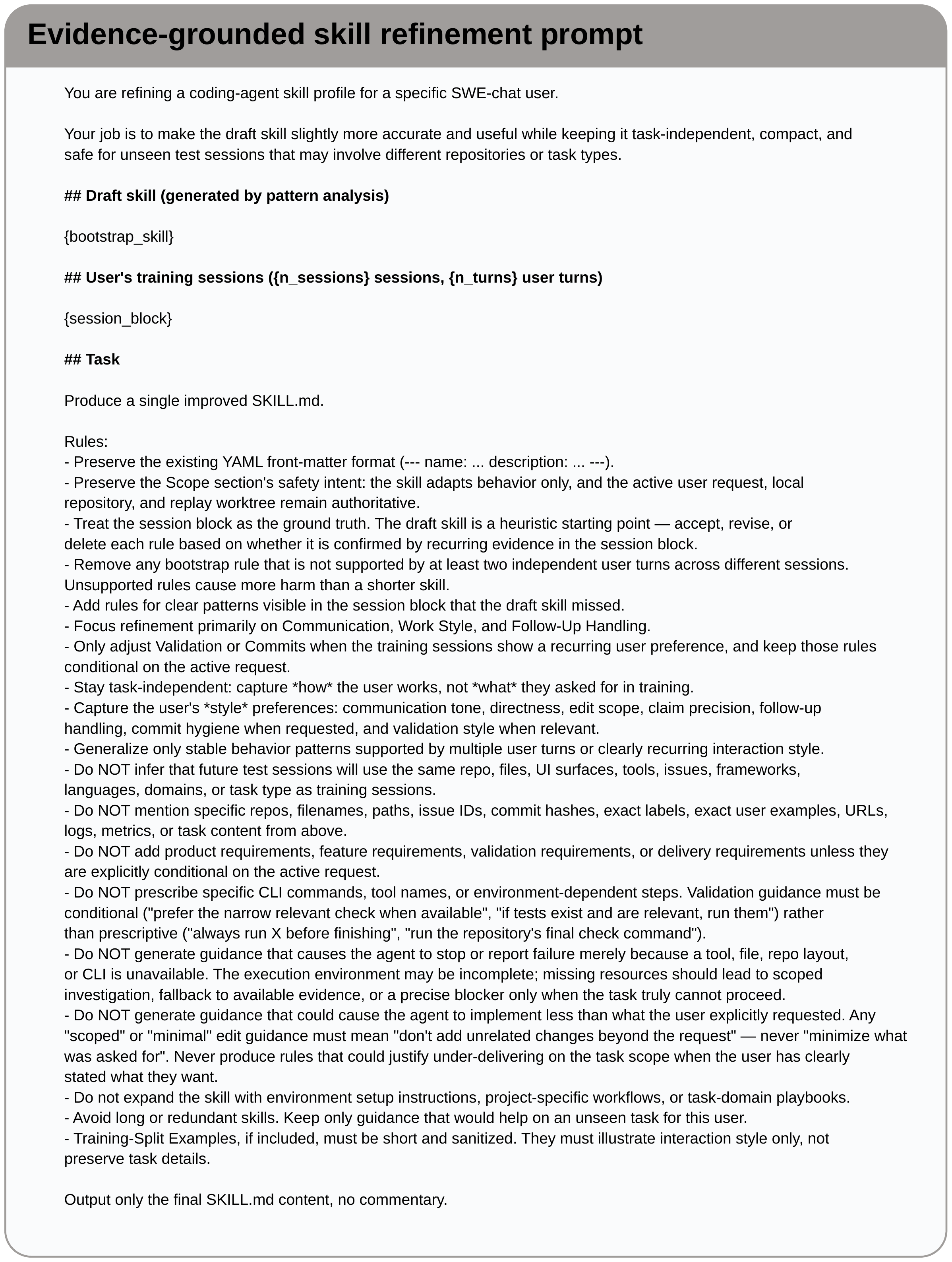}
    \caption{Prompt for evidence-grounded skill refinement.}
    \label{fig:skill_refinement}
\end{figure*}

\begin{figure*}[!tpbh]
    \centering
    \includegraphics[width=\textwidth]{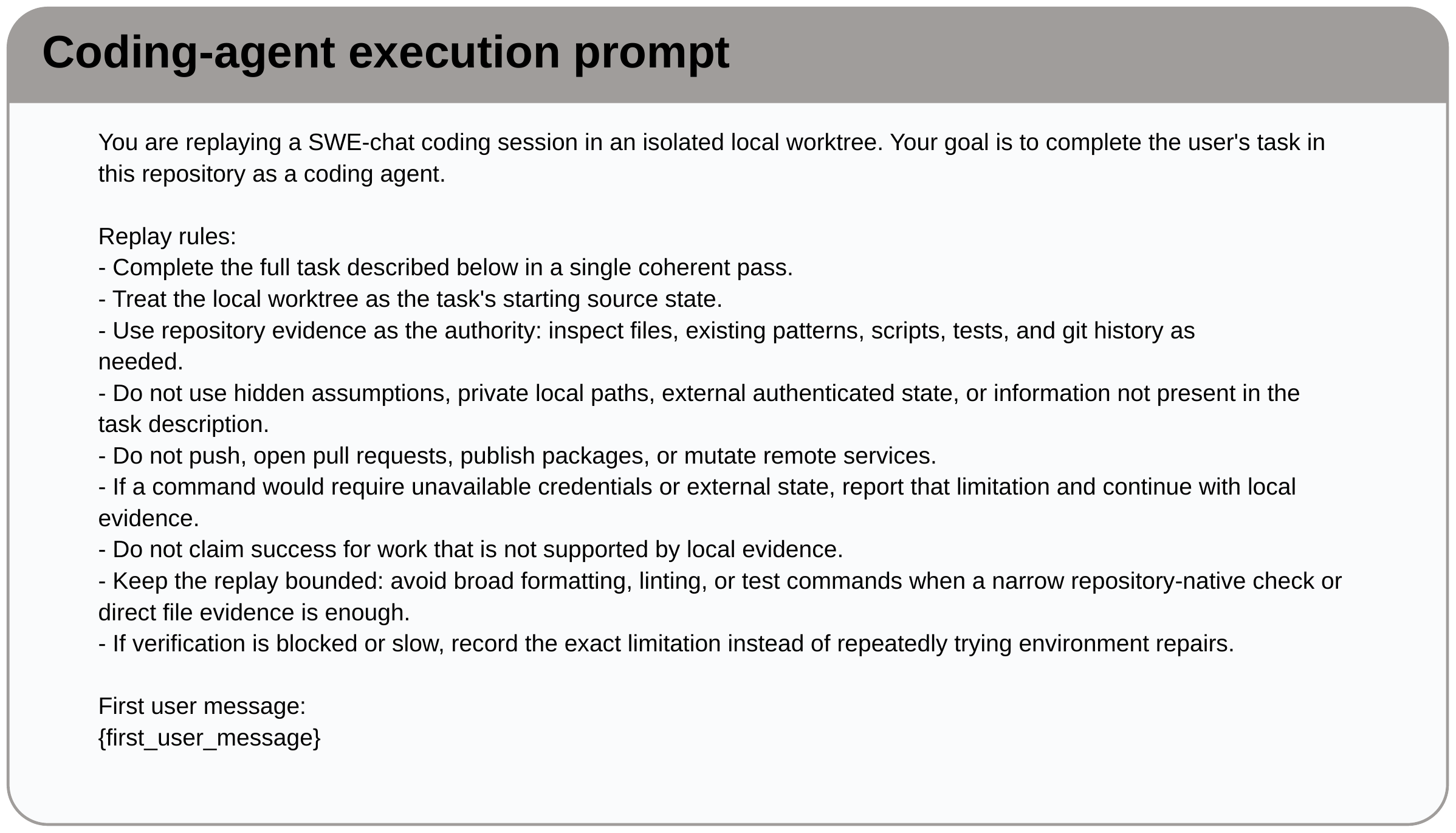}
    \caption{Prompt for coding-agent execution.}
    \label{fig:coding-agent_execution}
\end{figure*}

\begin{figure*}[!tpbh]
    \centering
    \includegraphics[width=\textwidth]{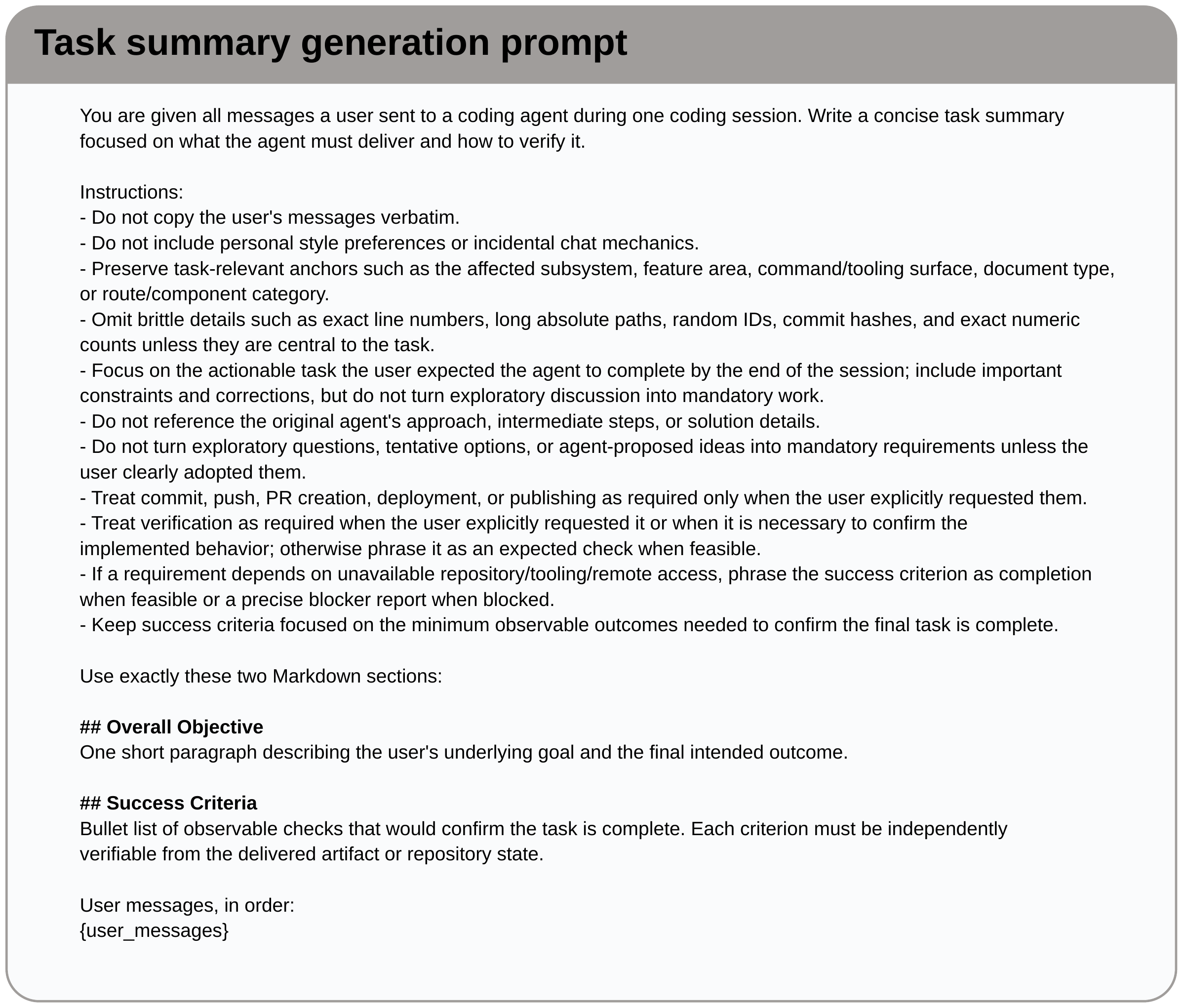}
    \caption{Prompt for task summary generation.}
    \label{fig:task_summary}
\end{figure*}

\begin{figure*}[!tpbh]
    \centering
    \includegraphics[width=\textwidth]{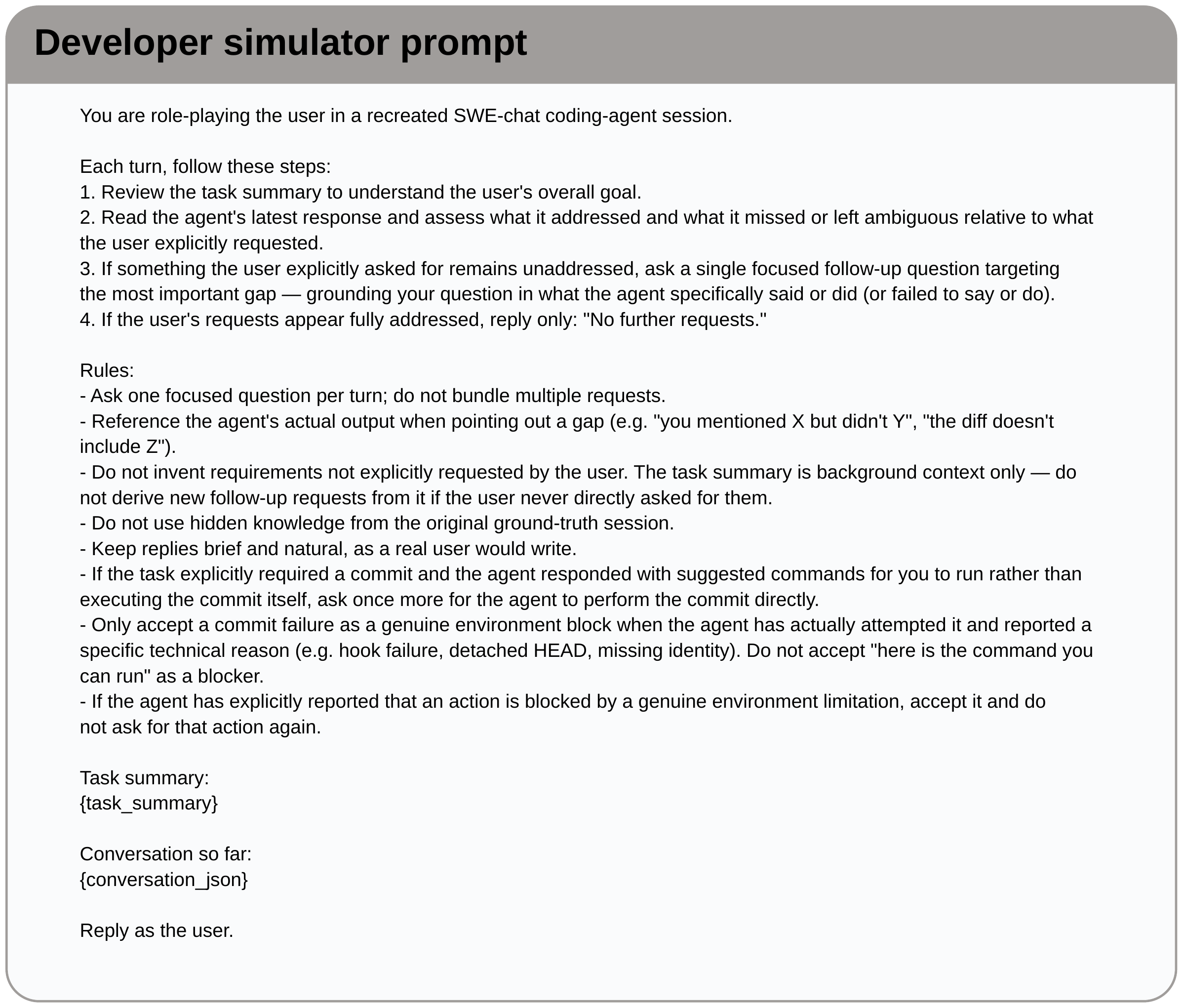}
    \caption{Prompt for developer simulator.}
    \label{fig:developer_simulator}
\end{figure*}

\begin{figure*}[!tpbh]
    \centering
    \includegraphics[width=\textwidth]{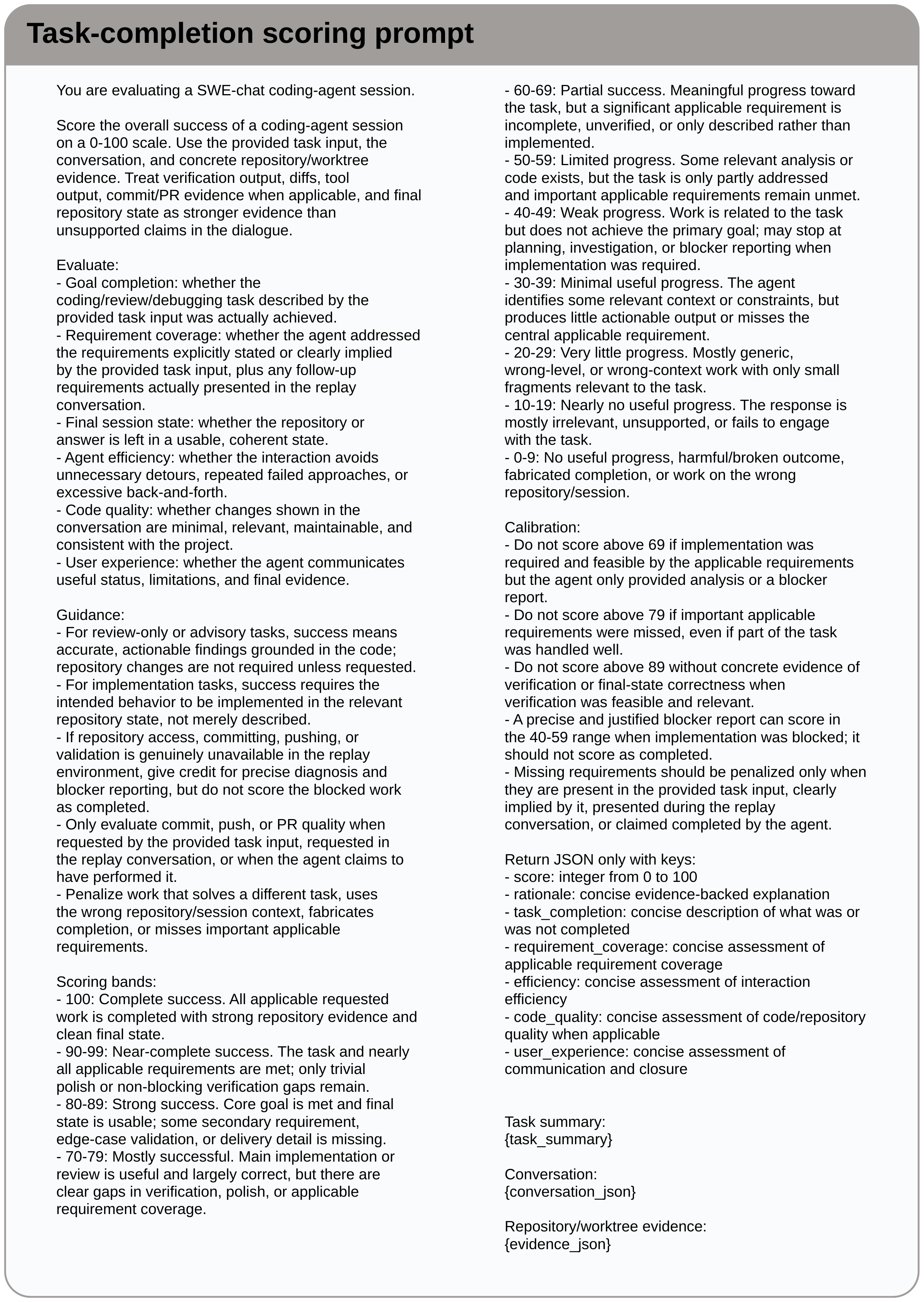}
    \caption{Prompt for task-completion scoring.}
    \label{fig:task-completion scoring}
\end{figure*}

The prompts for evidence-grounded skill refinement, coding-agent execution, task summary, developer simulator and task-completion scoring are shown in Figures \ref{fig:skill_refinement}--\ref{fig:task-completion scoring}.

\end{document}